\documentclass[aps,prl,preprint,tightenlines,superscriptaddress,showpacs,byrevtex]{revtex4}

\usepackage{graphicx}
\usepackage{dcolumn}
\usepackage{bm}
\usepackage{color}

\graphicspath{{eps/}}

\def\bbar {\overline{B}^{\,0}}

\def\dbar {\overline{D}^{\,0}}

\def\cp {$CP$}
\def\ra {\!\rightarrow\!}
\def\kbar {\overline{K}{}^{\,0}}
\def\dbar {\overline{D}{}^{\,0}}
\def\bbar {\overline{B}{}^{\,0}}
\def\rws {$R^{}_{WS}$}

\begin{document}

\preprint{\vbox{ \hbox{Belle Prerpint 2005-24}
                 \hbox{KEK Preprint 2005-35}
}}

\title{ \quad\\[0.5cm] \boldmath Measurement of the wrong-sign decays 
$D^0 \to K^+ \pi^- \pi^0$ and $D^0 \to K^+ \pi^- \pi^+\pi^-$,
and search for \cp\ violation}

\begin{abstract}

Using 281~fb$^{-1}$ of data from the Belle experiment 
recorded at or near the $\Upsilon$(4$S$) resonance, 
we have measured the rates of the ``wrong-sign'' decays 
$D^0 \ra K^+ \pi^-\pi^0$ and $D^0 \ra K^+ \pi^-\pi^+\pi^-$ 
relative to those of the Cabibbo-favored decays
$D^0 \ra K^- \pi^+\pi^0$ and $D^0 \ra K^- \pi^+\pi^+\pi^-$. 
These wrong-sign decays proceed via a doubly Cabibbo-suppressed 
amplitude or via $D^0$-$\dbar$ mixing; the latter has not 
yet been observed.
We obtain 
$R_\mathrm{WS}(K\pi\pi^0)=[\,0.229 \pm 0.015\,({\mathrm{stat.}})^{+0.013}_{-0.00
9}\,({\mathrm{sys.}})\,]\%$ and
$R_\mathrm{WS}(K3\pi)=[\,0.320 \pm 0.018\,({\mathrm{stat.}})^{+0.018}_{-0.013}\,
({\mathrm{sys.}})\,]\%$.
The \cp\ asymmetries are measured to be 
$-0.006\pm 0.053$ and $-0.018\pm0.044$ for the $K^+ \pi^- \pi^0$ 
and $K^+\pi^-\pi^+\pi^-$ final states, respectively.

\end{abstract}

\affiliation{Budker Institute of Nuclear Physics, Novosibirsk}
\affiliation{Chiba University, Chiba}
\affiliation{Chonnam National University, Kwangju}
\affiliation{University of Cincinnati, Cincinnati, Ohio 45221}
\affiliation{Gyeongsang National University, Chinju}
\affiliation{University of Hawaii, Honolulu, Hawaii 96822}
\affiliation{High Energy Accelerator Research Organization (KEK), Tsukuba}
\affiliation{Hiroshima Institute of Technology, Hiroshima}
\affiliation{Institute of High Energy Physics, Chinese Academy of Sciences, Beijing}
\affiliation{Institute of High Energy Physics, Vienna}
\affiliation{Institute for Theoretical and Experimental Physics, Moscow}
\affiliation{J. Stefan Institute, Ljubljana}
\affiliation{Kanagawa University, Yokohama}
\affiliation{Korea University, Seoul}
\affiliation{Kyungpook National University, Taegu}
\affiliation{Swiss Federal Institute of Technology of Lausanne, EPFL, Lausanne}
\affiliation{University of Ljubljana, Ljubljana}
\affiliation{University of Maribor, Maribor}
\affiliation{University of Melbourne, Victoria}
\affiliation{Nagoya University, Nagoya}
\affiliation{Nara Women's University, Nara}
\affiliation{National Central University, Chung-li}
\affiliation{National United University, Miao Li}
\affiliation{Department of Physics, National Taiwan University, Taipei}
\affiliation{H. Niewodniczanski Institute of Nuclear Physics, Krakow}
\affiliation{Nippon Dental University, Niigata}
\affiliation{Niigata University, Niigata}
\affiliation{Nova Gorica Polytechnic, Nova Gorica}
\affiliation{Osaka City University, Osaka}
\affiliation{Osaka University, Osaka}
\affiliation{Panjab University, Chandigarh}
\affiliation{Peking University, Beijing}
\affiliation{Princeton University, Princeton, New Jersey 08544}
\affiliation{Saga University, Saga}
\affiliation{University of Science and Technology of China, Hefei}
\affiliation{Seoul National University, Seoul}
\affiliation{Shinshu University, Nagano}
\affiliation{Sungkyunkwan University, Suwon}
\affiliation{University of Sydney, Sydney NSW}
\affiliation{Tata Institute of Fundamental Research, Bombay}
\affiliation{Toho University, Funabashi}
\affiliation{Tohoku Gakuin University, Tagajo}
\affiliation{Tohoku University, Sendai}
\affiliation{Department of Physics, University of Tokyo, Tokyo}
\affiliation{Tokyo Institute of Technology, Tokyo}
\affiliation{Tokyo Metropolitan University, Tokyo}
\affiliation{Tokyo University of Agriculture and Technology, Tokyo}
\affiliation{University of Tsukuba, Tsukuba}
\affiliation{Virginia Polytechnic Institute and State University, Blacksburg, Virginia 24061}
\affiliation{Yonsei University, Seoul}
   \author{X.~C.~Tian}\affiliation{Peking University, Beijing} 
   \author{Y.~Ban}\affiliation{Peking University, Beijing} 
   \author{K.~Abe}\affiliation{High Energy Accelerator Research Organization (KEK), Tsukuba} 
   \author{K.~Abe}\affiliation{Tohoku Gakuin University, Tagajo} 
   \author{H.~Aihara}\affiliation{Department of Physics, University of Tokyo, Tokyo} 
   \author{K.~Arinstein}\affiliation{Budker Institute of Nuclear Physics, Novosibirsk} 
   \author{Y.~Asano}\affiliation{University of Tsukuba, Tsukuba} 
   \author{V.~Aulchenko}\affiliation{Budker Institute of Nuclear Physics, Novosibirsk} 
   \author{T.~Aushev}\affiliation{Institute for Theoretical and Experimental Physics, Moscow} 
   \author{A.~M.~Bakich}\affiliation{University of Sydney, Sydney NSW} 
% ...
\collaboration{The Belle Collaboration}

\date{\textbf{\today}}

\maketitle

Studies of mixing in the $K^0$-$\kbar$ and $B^0$-$\bbar$ meson
systems~\cite{ref:Good} have had an important impact on the development
of the Standard Model (SM). The latter allowed the top quark mass to be
predicted prior to its direct observation.
In contrast, the $D^0$-$\dbar$ mixing rate is strongly 
suppressed by Cabibbo-Kobayashi-Maskawa (CKM) factors and the
GIM mechanism~\cite{ref:Glashow}; the SM predicted rate is far 
below current experimental upper limits. Observation of mixing 
significantly larger than this prediction could indicate new 
physics~\cite{ref:cicerone}. 
Previously, $D^0$-$\dbar$ mixing has been searched for 
in ``wrong-sign'' (WS) 
$D^0\ra K^+\pi^-$ decays~\cite{ref:e791,ref:others,ref:LiJin},
in WS $D^0\ra K^+\pi^-\pi^0$ and
$D^0\ra K^+\pi^-\pi^+\pi^-$ decays~\cite{ref:e791,cleok2p,cleok3p},
and in Dalitz-plot analyses of 
$D^0\ra K^0_s\pi^+\pi^-$ decays~\cite{ref:D02kspipi}.
Here we investigate the WS multi-body modes
$D^0\ra K^+ \pi^-(n\pi)$~\cite{ref:charge} with a data 
sample more than 30 times larger than that of previous 
studies. These modes
can arise from a $D^0$ mixing into $\dbar$ and subsequently
decaying via the ``right-sign''(RS) Cabibbo-favored (CF) decay
$\dbar\ra K^+ \pi^-(n\pi)$.  The final states 
can also arise from a doubly-Cabibbo-suppressed (DCS) amplitude; 
the ratio of DCS decays to CF decays can be used to measure the 
CKM phase $\phi_3$ in $B^+\rightarrow D^0K^+$~\cite{ads}.

In this Letter we present measurements of the ratio of
rates for WS to RS decays,
$R^{(K^+\pi^-\pi^0)}_\mathrm{WS}\equiv \Gamma(D^0\ra K^+\pi^-\pi^0)/\Gamma(D^0\ra K^-\pi^+\pi^0)$ and
$R^{(K^+\pi^-\pi^+\pi^-)}_\mathrm{WS}\equiv \Gamma(D^0\ra K^+\pi^-\pi^+\pi^-)/\Gamma(D^0\ra K^-\pi^+\pi^+\pi^-)$. 
Assuming negligible \cp\ violation, this ratio is
given by~\cite{ref:Blaylock}
\begin{eqnarray}
R^{}_{\rm WS}& = & R^{}_D + \sqrt{R^{}_D}\,y' +
\frac{1}{2}(x'^2+y'^2)\,,
\label{eqn:main}
\end{eqnarray}
where $R^{}_D$ is the ratio of the 
magnitudes squared of the DCS to CF amplitudes;
and $x'$ and $y'$ are ``rotated'' versions of 
the mixing parameters $x\equiv\Delta m/\overline{\Gamma}$ 
and $y\equiv\Delta\Gamma/2\overline{\Gamma}$:
$x'=x\cos\delta+y\sin\delta$ and $y'= y\cos\delta-x\sin\delta$, where 
$\delta$ is an effective strong phase difference between the
DCS and CF amplitudes~\cite{endnote}. The parameters $x$ and $y$
are mode-independent, depending only on the 
differences in mass ($\Delta m$) and decay width
($\Delta \Gamma$) between the two $D^0$-$\dbar$ mass 
eigenstates, and their mean decay width ($\overline{\Gamma}$). 


  The data sample consists of 281 $\mathrm {fb^{-1}}$ recorded by
  the Belle experiment at KEKB~\cite{KEKB}, an asymmetric
  $e^+e^-$ collider operating at or near the $\Upsilon$(4$S$)
  resonance. 
The Belle detector is a large-solid-angle 
magnetic spectrometer consisting of a silicon vertex detector (SVD), a 
50-layer central drift chamber (CDC), an array of aerogel threshold 
\v{C}erenkov counters (ACC), a barrel-like arrangement of time-of-flight 
scintillation counters (TOF), and an electromagnetic calorimeter (ECL),
all located inside a superconducting solenoid 
coil that provides a 1.5~T magnetic field.  An iron flux-return 
outside the coil is instrumented to detect $K_L^0$ mesons and to 
identify muons (KLM).  The detector is described in detail 
elsewhere~\cite{ref:Belle,ref:Ushiroda}. 

We consider the decay chain 
$D^{*+}\ra D^0\pi_s^+\ra K\pi (n\pi)\pi_s^+$, 
where the ``slow'' pion $\pi_s^+$ has a characteristic soft momentum 
spectrum. The charge of $\pi^{}_s$ is used to identify whether a 
$D^0$ or $\dbar$ was initially produced. We require that all tracks 
have at least two SVD hits in both $r$-$\phi$ and $z$ coordinates. 
We use information  from the TOF, ACC, and CDC  
to select kaons (pions) with momentum dependent efficiencies 
of $80 \textendash 95\%$ ($90 \textendash 95\%$) and 
pion (kaon) misidentification probabilities of 
$5 \textendash 20\%$ ($15 \textendash 20\%$).
To suppress background from semileptonic decays, 
we remove tracks identified as electrons (muons)
based on ECL (KLM) information. 
We select $\pi^0$ candidates that satisfy
$118$~MeV/$c^2<M^{}_{\gamma\gamma}<150$~MeV/$c^2$
($\pm3\sigma$ in resolution); we then apply a mass constrained fit
for the photons. We require photon energies to 
be larger than 60 (120)~MeV in the barrel (endcap) region.

$D^0\ra K^+\pi^-\pi^0$ candidates are reconstructed by 
combining two oppositely-charged tracks with a $\pi^0$ candidate 
having $p>310$~MeV/$c$ in the center-of-mass (CM) frame. The 
$K^+\pi^-\pi^0$ invariant mass is required to be in the 
range 1.78\textendash1.92~GeV/$c^2$ ($\pm6\sigma$ in resolution). 
To reject background from $D^0\ra K^-\pi^+\pi^0$ in which 
the $K$ is misidentified as $\pi$ and the $\pi$ 
as $K$, we calculate $m_{K\pi\pi^0}$ with 
the $K$ and $\pi$ assignments swapped and reject events 
having $m_{K\pi\pi^0({\rm{swapped}})}$ in the range 
1.78\textendash1.90 GeV$/c^2$. 

$D^0\ra K^+\pi^-\pi^+\pi^-$ candidates are formed 
from combinations of four charged tracks; $m^{}_{K3\pi}$ 
is required to be in the range 
1.81\textendash1.91 GeV/$c^2$ ($\pm7\sigma$).
To reject background due to misidentification of $D^0\rightarrow K^-\pi^+\pi^-\pi^+$,
we calculate $m^{}_{K3\pi}$ with the $K$ and $\pi$ 
assignments swapped and reject events satisfying
$|m^{}_{K3\pi({\rm{swapped}})}-m^{}_{D^0}|<$ 20 MeV$/c^2$. 
The Cabibbo-suppressed decay $D^0\ra\kbar K^+\pi^-$ 
followed by $\kbar\ra\pi^+\pi^-$ can also mimic the WS signal;
to reject this background, we calculate $m^{}_{\pi^+\pi^-}$ 
for both oppositely-charged pion combinations and reject events 
satisfying
$|m_{\pi^+\pi^-}- m^{}_{K^0} |<$ 16 MeV$/c^2$.

The charged $D^0$ daughters are required to originate from a 
common vertex. The $D^0$ momentum vector is extrapolated back 
to the interaction point (IP) profile and a production vertex 
is determined. 
The $D^{*+}$ candidate is  
then formed by combining the $D^0$ candidate with a $\pi_s^+$. 
We refit the $\pi^+_s$ track, requiring that it intersect
the $D^0$ production point; this greatly suppresses combinatorial 
background and improves the resolution 
on the energy released in the $D^*$ decay,
$Q \equiv M_{\pi_s^+K^+\pi^- 
(n\pi)}-M_{K^+\pi^- 
(n\pi)}-m_{\pi^+_s}$.
For $D^{*\,+}\ra D^0\pi^+_s$ decays, $Q$ is only 5.85~MeV
(slightly above threshold) and provides substantial background
rejection. We subsequently require $Q\!<\!12$~MeV,
which is $>$\,99\% efficient.

To eliminate $D$ mesons produced in $B\overline{B}$ events and 
further suppress combinatorial background, 
the reconstructed $D^{*+}$ momentum in the CM 
frame is required to be greater than 2.5~GeV/$c$.
Finally, we require that the $\chi^2$ per degree of freedom (d.o.f) 
resulting from the $D^0$ vertex fit, the IP vertex fit, and the 
$\pi_s$ track refit be satisfactory. The fraction of events 
containing multiple signal candidates is less than 3\% for both modes (and 
is the same for RS and WS decays);
multiple signal candidates are retained 
for subsequent analysis.

We determine the RS and WS signal yields by performing 
binned maximum likelihood fits in $M$-$Q$ space with 
$M = M_{K\pi\,(n\pi)}$.    
The signal and background distributions are 
determined using a large Monte Carlo (MC) sample~\cite{ref:Monte carlo}. 
The backgrounds can be divided into three categories: 
$(a)$ ``random $\pi_s$'' background, 
in which a random $\pi^+$ is combined with a true 
$\dbar\ra K^+\pi^- (n\pi)$ decay; 
$(b)$ charm decay background other than (a); and
$(c)$ background from continuum
$e^+e^-\ra u\bar u, d\bar d$, or $s\bar s$ production. 

The RS signal shape as predicted by MC simulation is 
parameterized in $M$ with a sum of a double Gaussian and a double bifurcated 
Gaussian with common mean, and in $Q$ with a bifurcated Student's $t$ 
function. Background distributions are parameterized with 
similar empirical expressions determined from MC simulation. 
In the RS sample fit, the mean and width of the signal 
distribution are left free to vary, while
other parameters are fixed to MC values. The relative normalizations of 
individual background categories are fixed to MC values for 
the $D^0\ra K^+\pi^-\pi^0$ fit, and left free for 
the $D^0\ra K^+\pi^-\pi^+\pi^-$ fit. 
In the WS sample fit, the mean and width of the signal are fixed 
to the values obtained from the RS fit; the normalizations of 
the backgrounds are left free to vary.

\begin{figure}
\begin{center}
\includegraphics[width=0.45\textwidth]{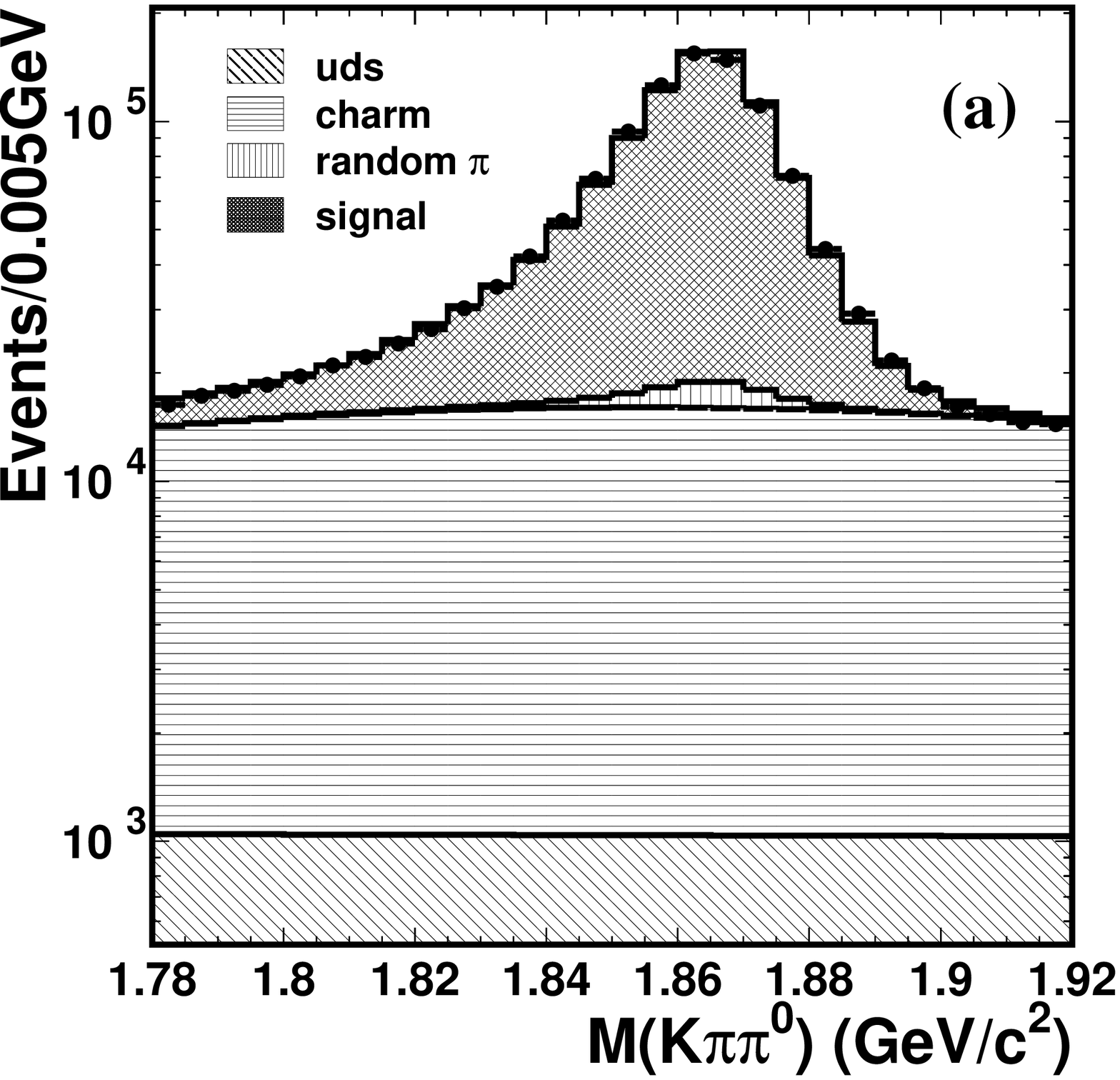}%
\includegraphics[width=0.45\textwidth]{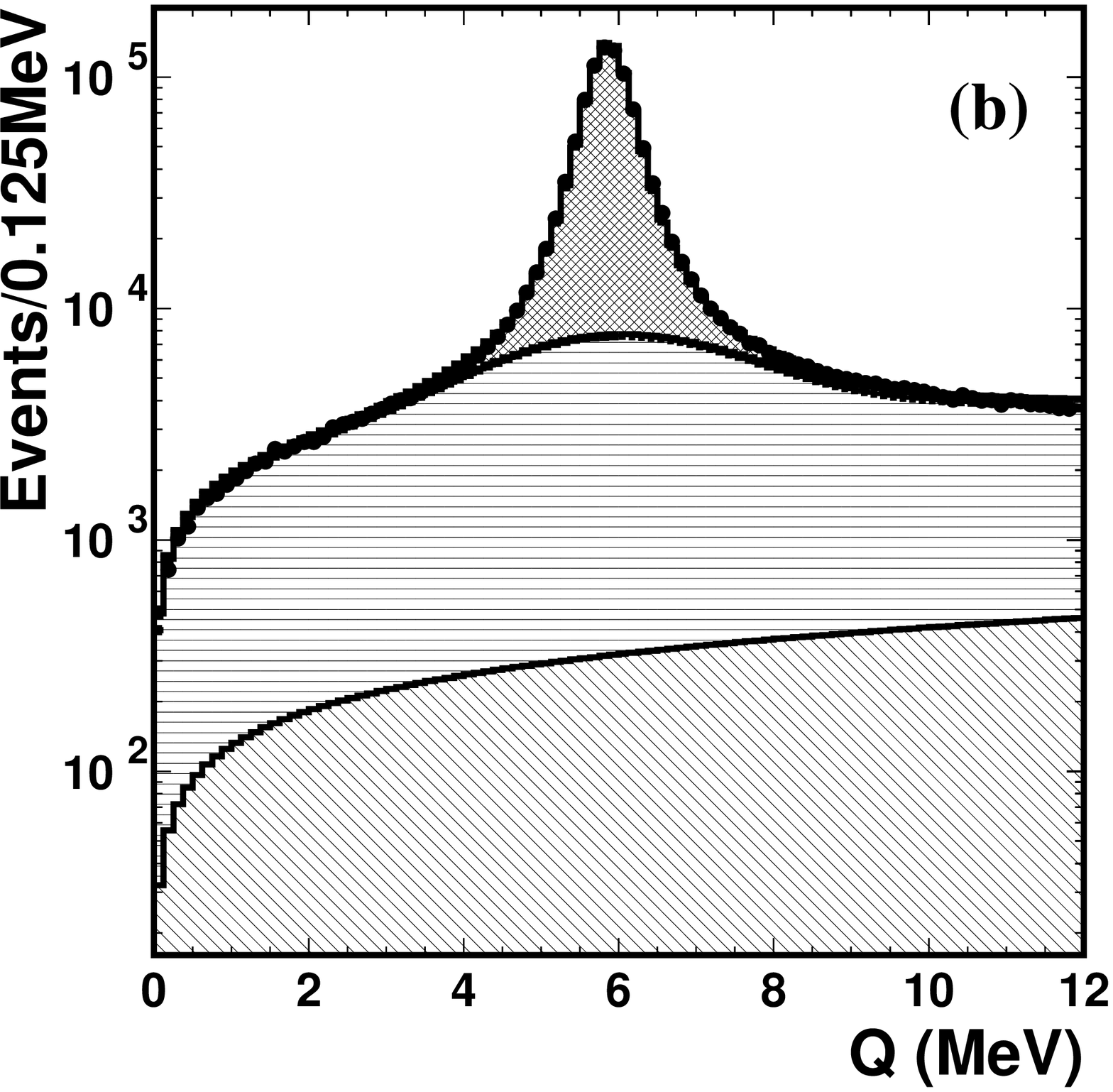}\break
\includegraphics[width=0.45\textwidth]{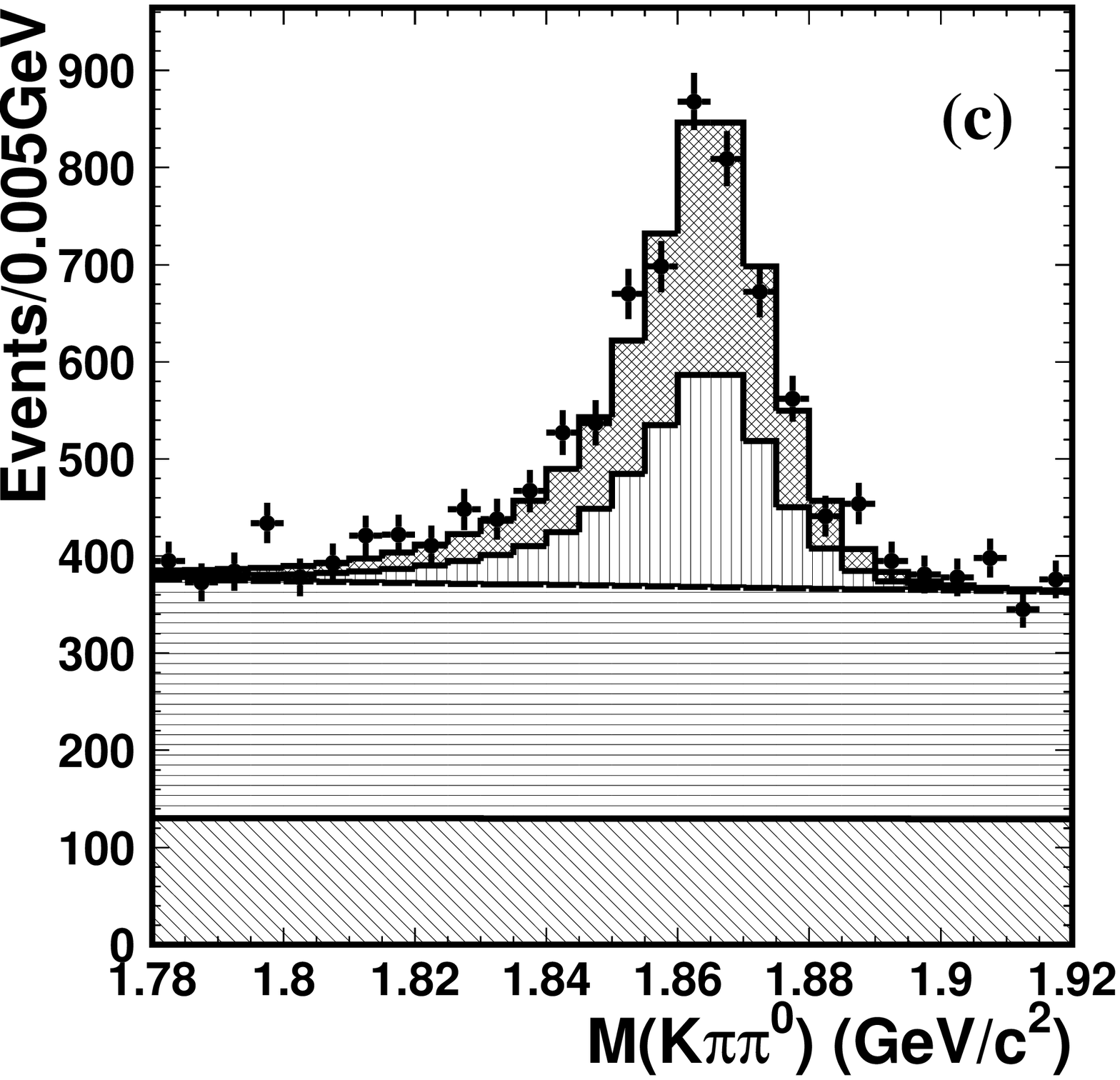}%
\includegraphics[width=0.45\textwidth]{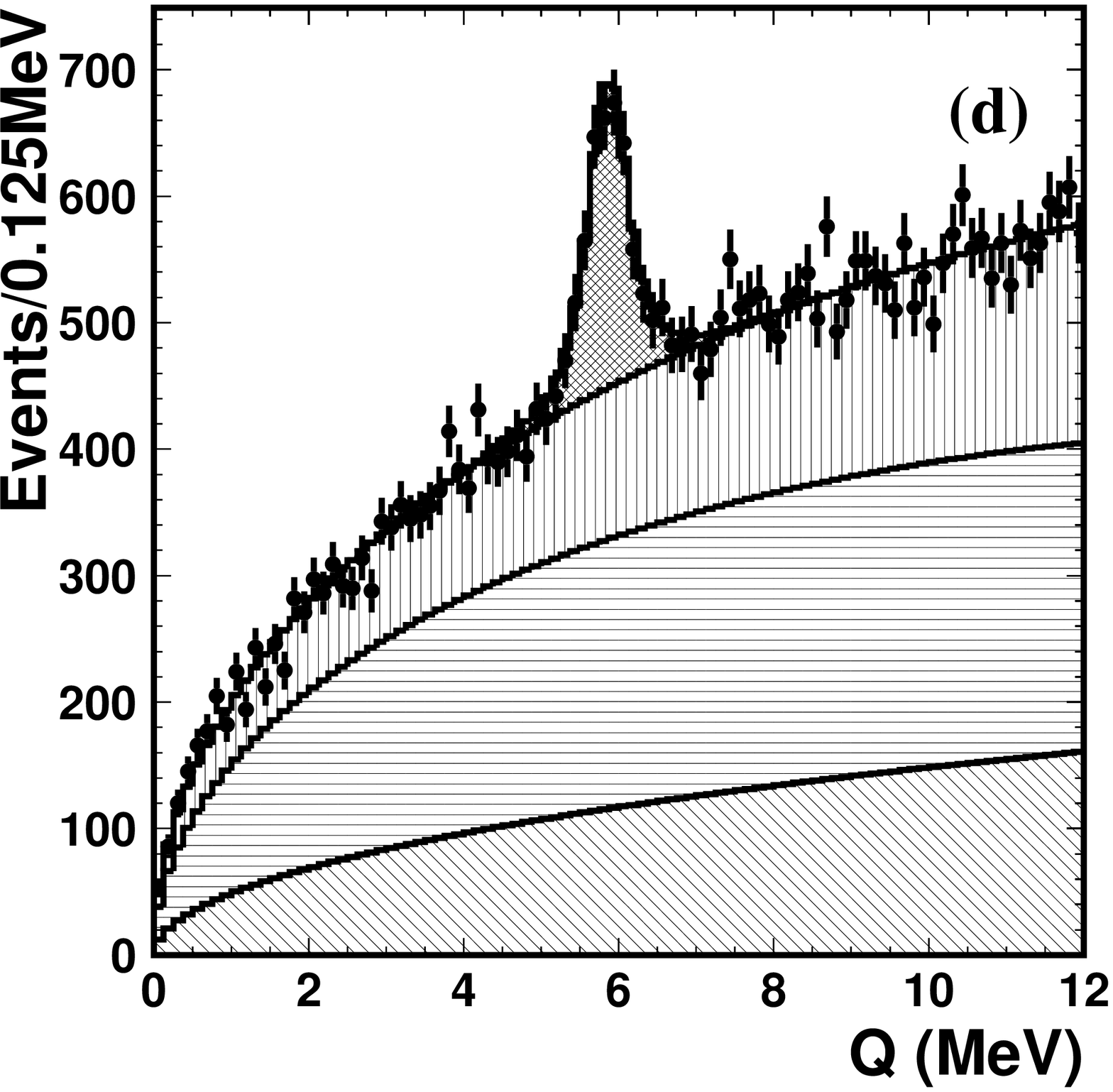}
\caption{\label{k2p} 
Results of the $M$-$Q$ fit for
$D^0\rightarrow K^+\pi^-\pi^0$, in projections onto (a) RS 
$M_{K\pi\pi^0}$ with 
$0~{\rm MeV}\!< Q < 12.0\,{\mathrm{MeV}}$; 
(b) RS $Q$ with $1.780\,{\mathrm {GeV}}/c^2 < M_{K\pi\pi^0} < 1.920\,{\mathrm{GeV}}/c^2$; 
(c) WS $M_{K\pi\pi^0}$ with $5.31\,{\mathrm{MeV}} < Q < 6.42 \,{\mathrm{MeV}}$; 
and (d) WS Q with $1.844 \,\,{\mathrm{GeV}}/c^2 \,< \,M_{K\pi\pi^0} \,< \,1.887\,\, {\mathrm{GeV}}/c^2$.
}
\end{center}
\end{figure}

\begin{figure}
\begin{center}
\includegraphics[width=0.45\textwidth]{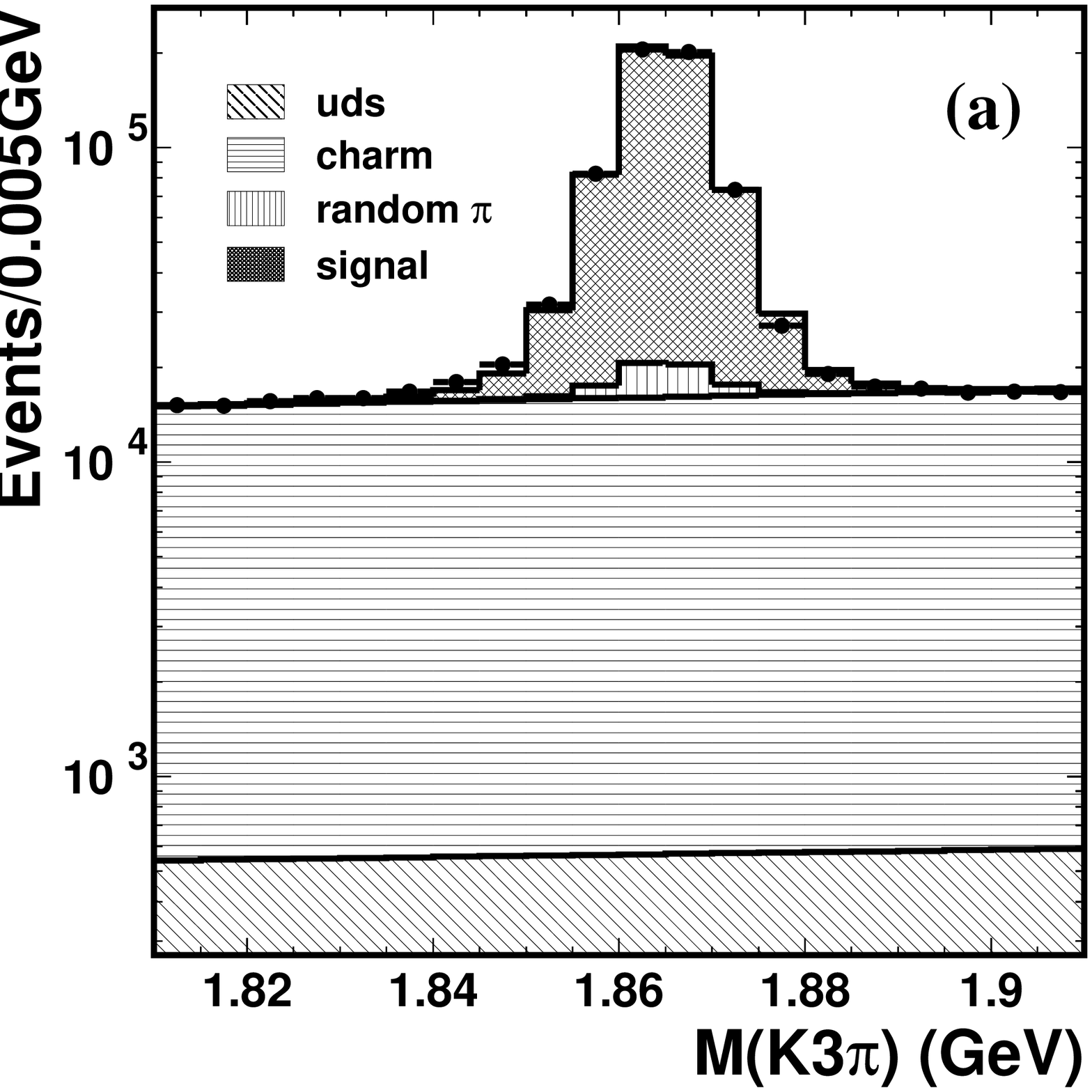}%
\includegraphics[width=0.45\textwidth]{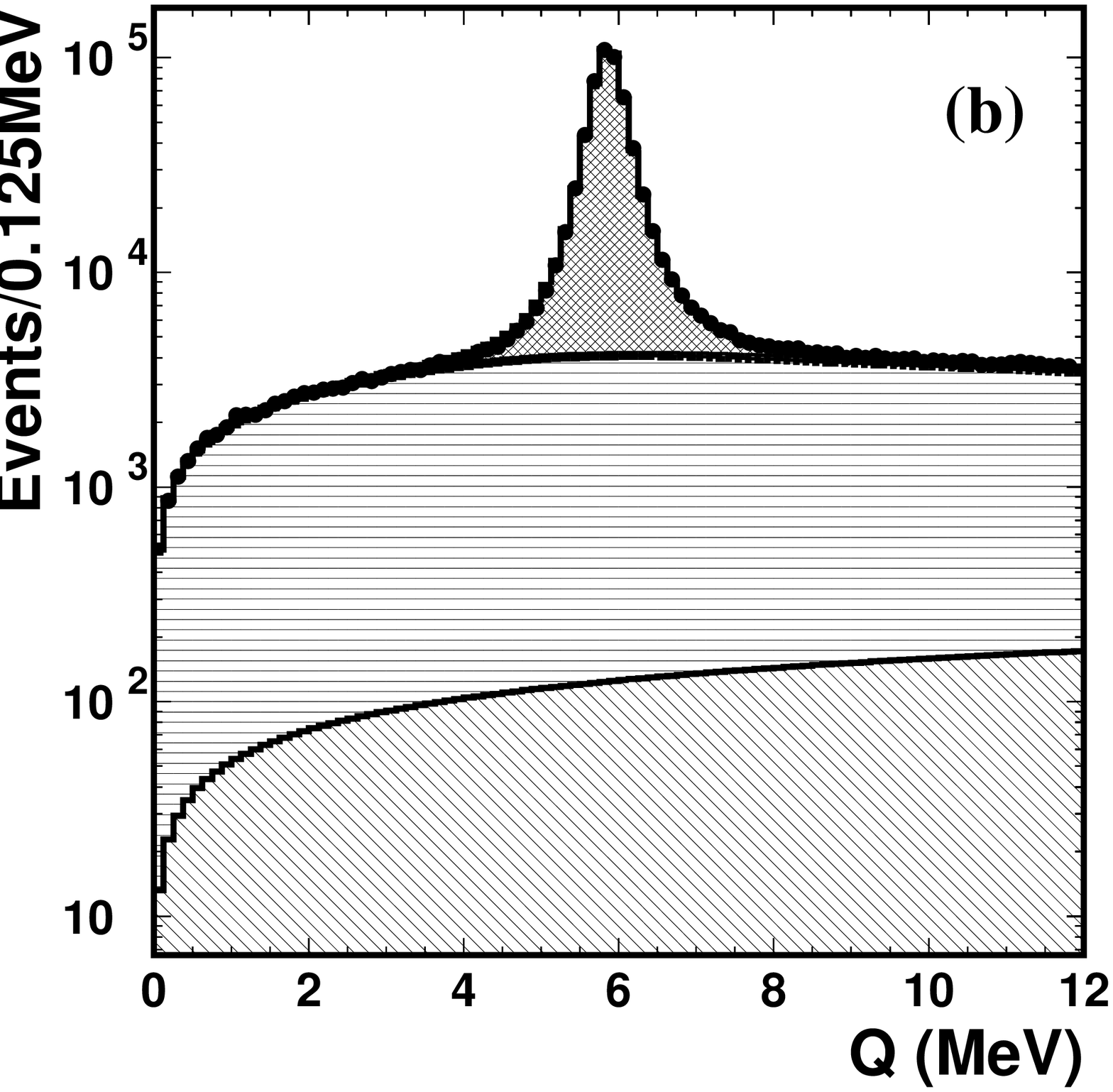}\break
\includegraphics[width=0.45\textwidth]{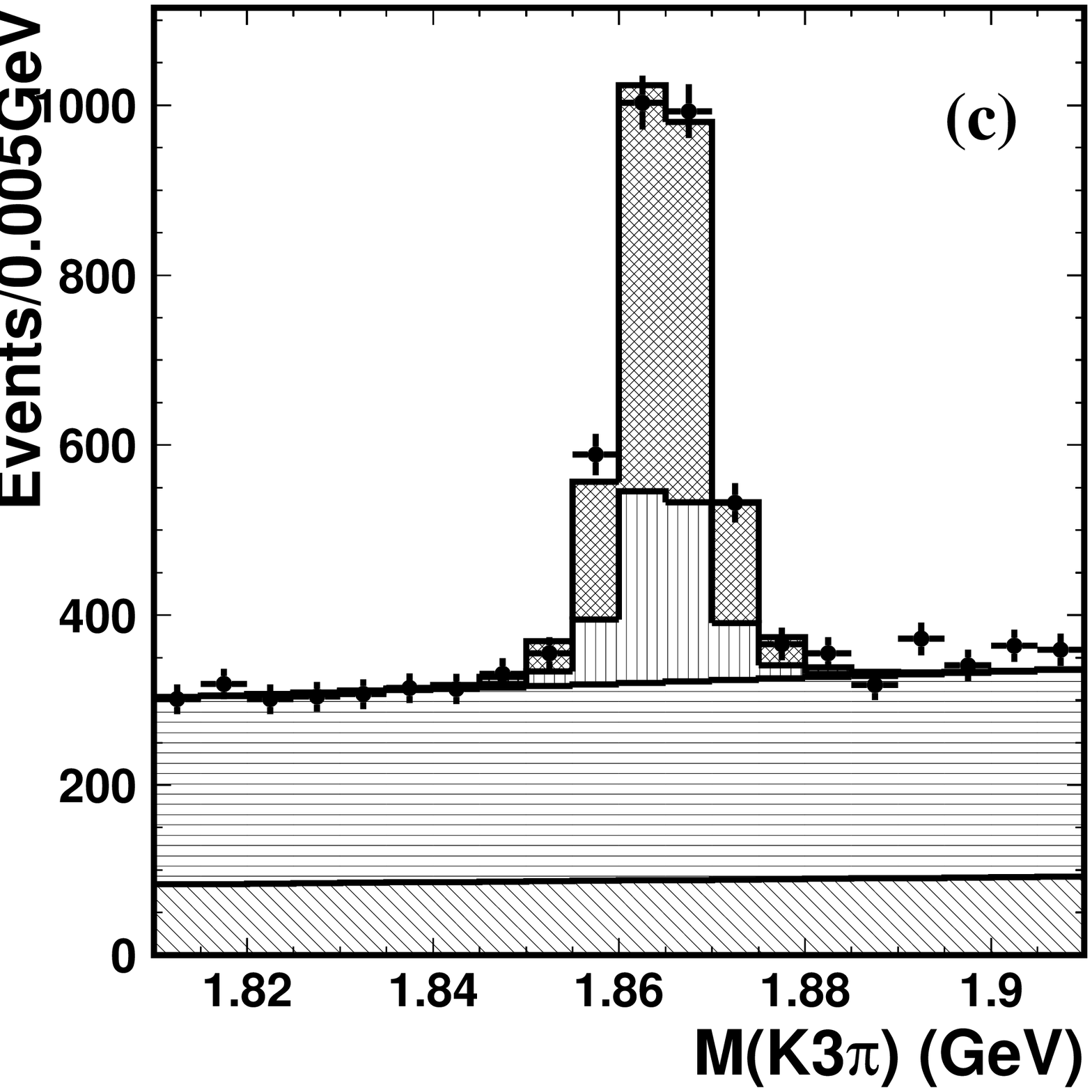}%
\includegraphics[width=0.45\textwidth]{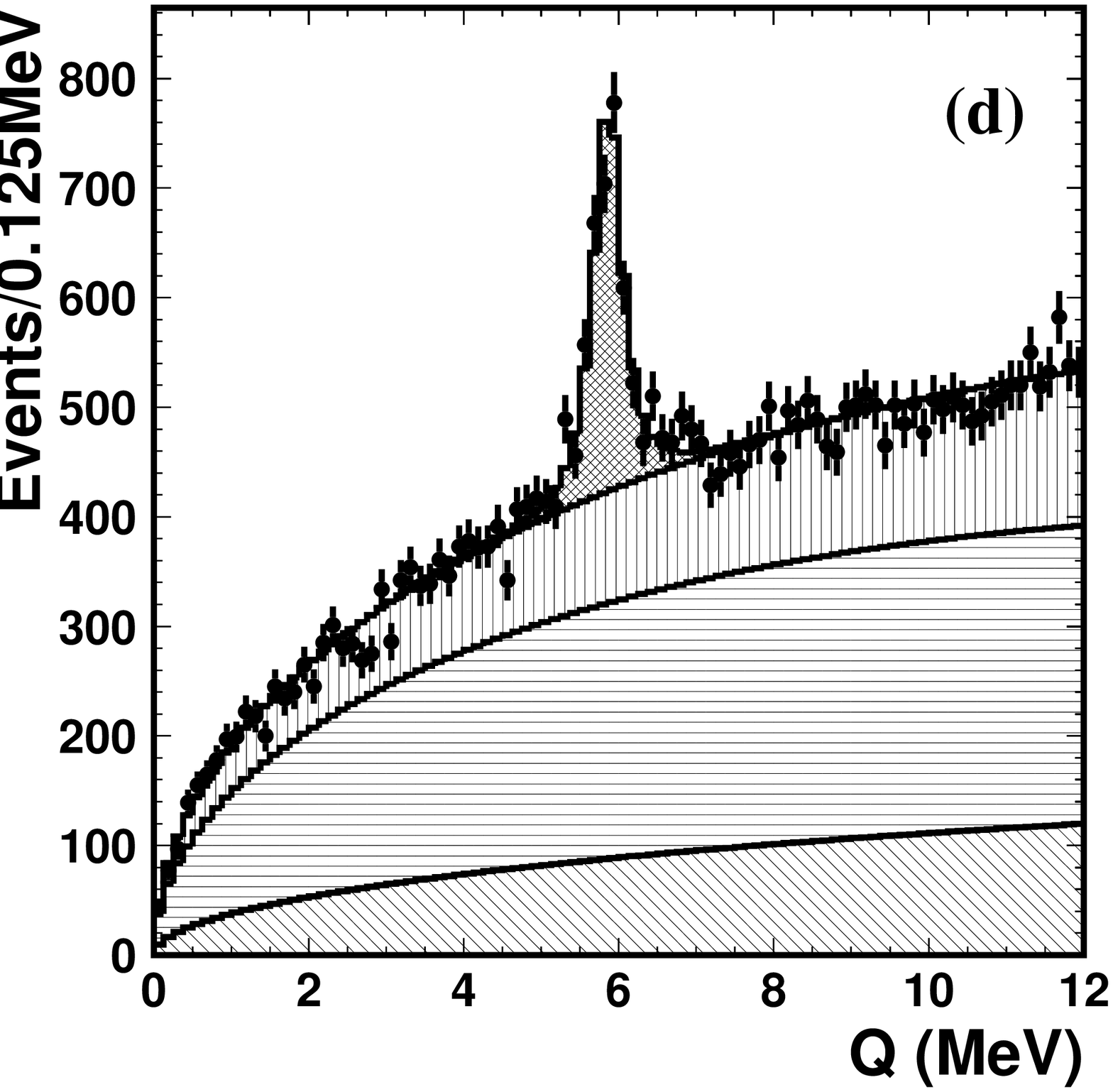}
\caption{\label{k3p} 
Results of the $M$-$Q$ fit for
$D^0\rightarrow K^+\pi^-\pi^+\pi^- $, in projections onto 
(a) RS $M_{K3\pi}$ with 
$0~{\rm MeV}\,<\,Q\,<\,12.0\,{\mathrm{MeV}}$; 
(b) RS $Q$ with $1.810\,{\mathrm{GeV}}/c^2 < M_{K3\pi} < 1.910 \,{\mathrm{GeV}}/c^2$; 
(c) WS $M_{K3\pi}$ with $5.47\,{\mathrm{MeV}} < Q < 6.28\,{\mathrm{MeV}}$; 
and (d) WS Q with $1.852\,{\mathrm{GeV}}/c^2 < M_{K3\pi} < 1.878 \,{\mathrm{GeV}}/c^2$.
}
\end{center}
\end{figure}

The RS sample fit obtains a signal yield of 
$(8.683 \pm 0.002) \times 10^{5}$ for $D^0\ra K^-\pi^+\pi^0$ and 
$(5.259 \pm 0.002 )\times 10^{5}$ for $D^0\ra K^-\pi^+\pi^+\pi^-$. 
The WS fit finds $1978\pm 104$ 
for $D^0\ra K^+\pi^-\pi^0$ and $1721\pm 75$ 
for $D^0\ra K^+\pi^-\pi^+\pi^-$. 
The fit results are projected onto the $M$ and $Q$ 
distributions in Fig.~\ref{k2p} for $D^0\ra K^+\pi^-\pi^0$ 
and in Fig.~\ref{k3p} for $D^0\ra K^+\pi^-\pi^+\pi^-$. 
The hatched histograms show the fit results and the
points with error bars show the data.

In $D^0\rightarrow K\pi(n\pi)$ decays, 
intermediate resonances dominate the decay rate and cause
a nonuniform event distribution in phase space. Since RS and WS 
decays may have different resonant substructure, their acceptances 
may differ. We correct the event yields for acceptance and
reconstruction efficiency as follows.
For $D^0\ra K^\pm\pi^\mp\pi^0$, we determine efficiencies using
MC simulation in bins of ($M^2_{K\pi}$, $M^2_{\pi\pi^0}$);
for $D^0\ra K^\pm\pi^\mp\pi^+\pi^-$, we use bins in 
a five-dimensional space comprised of the invariant mass squared
for various $K,\pi$ combinations. We then calculate 
efficiency-corrected signal yields in each bin for 
the RS and WS samples. The background is taken to 
be the overall background yield multiplied by the 
fraction falling in that bin; the distribution 
of background among the bins is taken from the 
sideband $|Q-5.85\,\,{\rm MeV}|>$ 2.0~MeV.
The resulting signal yields are summed over 
all bins, and the ratio of the total signal 
yields gives $R^{}_{\rm WS}$. The results are
$R^{K^+\pi^-\pi^0}_\mathrm{WS} =
(2.29\,\pm\,0.15) \times 10^{-3}$ and
$R^{K^+\pi^-\pi^+\pi^-}_\mathrm{WS} =
(3.20\,\pm\,0.18)\times 10^{-3}$, where the
errors are statistical only.

The average efficiency for a mode is obtained by
dividing the signal yield from the $M$-$Q$ fit by 
the total efficiency-corrected signal yield; the 
ratio of average efficiencies 
$\langle\varepsilon^{}_{\rm RS}\rangle /
\langle\varepsilon^{}_{\rm WS}\rangle$
is $1.01 \pm 0.05$ for $D^0\ra K^\pm\pi^\mp\pi^0$ and  
$0.98 \pm 0.04$ for $D^0\ra K^\pm\pi^\mp\pi^+\pi^-$.


Contributions to the systematic uncertainty on \rws\ are listed in 
Table~\ref{systematics}: the size of each term is assessed 
by varying the analysis as described below and repeating the fits. 
Many effects cancel in the ratio due to the similar kinematics
of the RS and WS modes; one distinction is the
significant background contribution to the WS sample. 
We vary the selection criteria over reasonable ranges 
(the WS yield changes by $\sim$\,10\%); the largest 
positive and negative variations in 
$R^{}_{\rm WS}$ are assigned as systematic errors.
We check the parameterization of the signal shape
by varying the means and widths in $M$ and $Q$ by~$\pm 1\sigma$. 
We check background fractions and parameterizations
by varying individual fractions and distribution 
parameters by $\pm 1 \sigma$;
we also try alternative functional forms.
We investigate possible fit bias by fitting a large 
MC RS sample; the small difference between the 
fitted yield and the true number of RS events is taken 
as an additional systematic error.
The total systematic error is obtained by combining the 
individual terms in quadrature.

\begin{table}[tb]
\begin{center}
\caption{Systematic uncertainties for $R^{}_{\rm WS}$, in 
percentage.}
\renewcommand{\arraystretch}{1.2}
\begin{tabular}{lcccc}
\hline
\hline
Source  & \multicolumn{2}{c}
{\hspace*{0.10in}$D^0\ra K^+\pi^-\pi^0$\hspace*{0.10in}} 
        & \multicolumn{2}{c}
{\hspace*{0.10in}$D^0\ra K^+\pi^-\pi^+\pi^-$\hspace*{0.10in}} \\
\hline 
Selection criteria &  $\ \ \ +5.22$ & $-2.38$ &  
$\ \ \ +5.25$ & $-3.78$  \\
Signal shape param.  &  $\ \ \ +0.09$ & $-0.10$ &  
$\ \ \ +0.10$ & $-0.10$  \\
Background fraction   &  $\ \ \ +0.00$ & $-0.07$ &  
$\ \ \ +0.01$ & $-0.01$  \\
Background param. &  $\ \ \ +0.42$ & $-2.89$ 
&\ \ \ $+0.34$ & $-0.59$ \\
Possible fit bias &  $\ \ \ +2.23$ & $-0.94$ &  $\ \ \ +0.91$ & $-0.88$  \\
\hline
Total     &  $\ \ \ +5.7$ & $-3.9$  &  $\ \ \ +5.4$  & $-4.0$   \\
\hline 
\hline
\end{tabular}
\label{systematics}
\end{center}
\end{table}

\begin{figure}
\begin{center}
\includegraphics[width=0.75\textwidth]{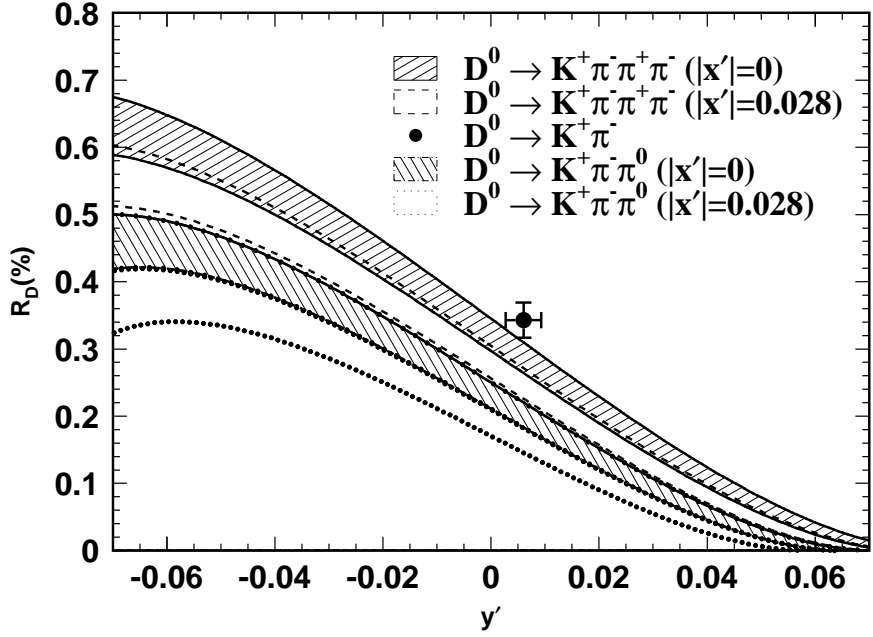}
\caption{
68.3\% CL bands for $R^{}_D$ as a function
of $y'$ for $x'=0$ and $|x'|=0.028$. The latter value
is the upper limit obtained from our analysis of
$D^0\ra K^+\pi^-$ decays assuming no \cp\ violation~\cite{ref:LiJin}.
The point with $1\sigma$ error bars is the result
from the $D^0\ra K^+\pi^-$ analysis for $x'\!=\!0$
(and no \cp\ violation). Note that $\delta$ and thus
$x',\,y'$ may differ for the three modes.
\label{fig:rd_y} }
\end{center}
\end{figure}

Assuming a value for $x'$, Eq.~(\ref{eqn:main}) can be used to
constrain $R^{}_D$ as a function of $y'$. This constraint is shown
in Fig.~\ref{fig:rd_y} for $x'\!=\!0$ and $|x'|\!=\!0.028$; the latter 
value is the 95\% CL upper limit on $|x'|$ obtained from our
previous analysis of $D^0\ra K^+\pi^-$ decays~\cite{ref:LiJin}.
Values of $(x',y')$ for different decay modes would 
be equivalent
if the strong phase differences ($\delta$) for the 
modes were equal. In the absence of mixing
(i.e., $x\!=\!y\!=\!0$), our measurements give
$R^{}_D(K\pi\pi^0)=(0.85\,^{+0.08}_{-0.07})\tan^4\theta^{}_C$ and 
$R^{}_D(K3\pi)=(1.18\,^{+0.10}_{-0.09})\tan^4\theta^{}_C$
($\theta^{}_C$ is the Cabibbo angle),
consistent with theoretical expectations~\cite{ref:cicerone2}.

By separately fitting the $D^0$ and $\dbar$ samples, we
measure the \cp\ asymmetry 
\begin{eqnarray*}
A^{}_{CP} & \!=\! & 
\frac{R_\mathrm{WS}^{D^0\rightarrow K^+\pi^-\,(n\pi)}-
     R_\mathrm{WS}^{\dbar\rightarrow K^-\pi^+\,(n\pi)}}
    {R_\mathrm{WS}^{D^0\rightarrow K^+\pi^-\,(n\pi)}+
     R_\mathrm{WS}^{\dbar\rightarrow K^-\pi^+\,(n\pi)}}\,. 
\end{eqnarray*}
We obtain 
$A^{}_{CP}(K\pi\pi^0)= -0.006 \pm 0.053$ and  
$A^{}_{CP}(K3\pi)= -0.018 \pm 0.044$, 
which are both consistent with zero. 
The systematic uncertainties are $<\!0.01$ 
(much smaller than the statistical errors) and are neglected.
The first value represents a large improvement 
over the previously-published result~\cite{cleok2p}; 
the second value has not been previously measured.

In summary, using 281 $\mathrm {fb^{-1}}$ of data 
we measure the ratio of WS to RS decay rates 
for $D^0\ra K^\pm\pi^\mp\pi^0$ and $D^0\ra K^\pm\pi^\mp\pi^+\pi^-$
to be
\begin{eqnarray*}
R^{K^+\pi^-\pi^0}_\mathrm{WS} & = & 
\left[\,2.29\,\pm\,0.15\,{\rm (stat)}^{\,+0.13}_{\,-0.09}\,{\rm (syst)}\,\right] \times 10^{-3}\\
 & & \\
R^{K^+\pi^-\pi^+\pi^-}_\mathrm{WS} & = & 
\left[\,3.20\,\pm\,0.18\,{\rm (stat)}^{\,+0.18}_{\,-0.13}\,{\rm (syst)}\,\right] \times 10^{-3}\,.\\
\end{eqnarray*} 
These results are 
much more precise than
previously-published results~\cite{ref:e791,cleok2p,cleok3p}.
The \cp\ asymmetries measured are consistent with zero.

\begin{acknowledgments}
We thank the KEKB group for the excellent operation of the
accelerator, the KEK cryogenics group for the efficient
operation of the solenoid, and the KEK computer group and
the NII for valuable computing and Super-SINET network
support.  We acknowledge support from MEXT and JSPS (Japan);
ARC and DEST (Australia); NSFC (contract No.~10175071, China); 
DST (India); the BK21 program of MOEHRD and the CHEP SRC 
program of KOSEF (Korea); KBN (contract No.~2P03B 01324, Poland); 
MIST (Russia); MHEST (Slovenia);  SNSF (Switzerland); 
NSC and MOE (Taiwan); and DOE~(USA).

\end{acknowledgments}


\begin{thebibliography}{99}
\bibitem{ref:charge}
Charge-conjugate states are implied throughout this paper.
We write $D^0\ra K^+ \pi^- (n\pi)$ to denote both 
$D^0\ra K^+\pi^-\pi^0$ ($n\!=\!1$) and 
$D^0\ra K^+\pi^-\pi^+\pi^-$ ($n\!=\!2$).
\end{thebibliography}
\end{document}